\documentclass[sigconf]{acmart}

\usepackage{booktabs} 
\usepackage{algorithm}
\usepackage{algpseudocode}

\begin{document}

\copyrightyear{2018} 
\acmYear{2018} 
\setcopyright{acmlicensed}
\acmConference[SIGSPATIAL '18]{26th ACM SIGSPATIAL International Conference on Advances in Geographic Information Systems}{November 6--9, 2018}{Seattle, WA, USA}
\acmBooktitle{26th ACM SIGSPATIAL International Conference on Advances in Geographic Information Systems (SIGSPATIAL '18), November 6--9, 2018, Seattle, WA, USA}
\acmPrice{15.00}
\acmDOI{10.1145/3274895.3274982}
\acmISBN{978-1-4503-5889-7/18/11}

\title{Creating Full Individual-level Location Timelines from Sparse Social Media Data}


\author{Nabeel Abdur Rehman}
\affiliation{%
  \institution{New York University}
}
\email{nabeel@nyu.edu}

\author{Kunal Relia}
\affiliation{%
  \institution{New York University}
}
\email{krelia@nyu.edu}

\author{Rumi Chunara}
\affiliation{%
  \institution{New York University}
}
\email{rumi.chunara@nyu.edu}


\begin{abstract}
In many domain applications, a continuous timeline of human locations is critical; for example for understanding possible locations where a disease may spread, or the flow of traffic. While data sources such as GPS trackers or Call Data Records are temporally-rich, they are expensive, often not publicly available or garnered only in select locations, restricting their wide use. Conversely, geo-located social media data are publicly and freely available, but present challenges especially for full timeline inference due to their sparse nature. We propose a stochastic framework, Intermediate Location Computing (ILC) which uses prior knowledge about human mobility patterns to predict every missing location from an individual's social media timeline. We compare ILC with a state-of-the-art RNN baseline as well as methods that are optimized for next-location prediction only. For three major cities, ILC predicts the top 1 location for all missing locations in a timeline, at 1 and 2-hour resolution, with up to 77.2\% accuracy (up to 6\% better accuracy than all compared methods). Specifically, ILC also outperforms the RNN in settings of low data; both cases of very small number of users (under 50), as well as settings with more users, but with sparser timelines. In general, the RNN model needs a higher number of users to achieve the same performance as ILC. Overall, this work illustrates the tradeoff between prior knowledge of heuristics and more data, for an important societal problem of filling in entire timelines using freely available, but sparse social media data.

\end{abstract}

%
%



\maketitle

\section{Introduction}

Using full location timelines (all locations an individual has been to) is essential to many societal applications including transportation management \cite{li2017urban,53}, urban sensing \cite{109}, event detection \cite{aoki2017early} and infectious disease dynamics \cite{32}. Combined with additional information about individuals, location timelines have been used to predict depressive moods \cite{107}, point-of-interest and location recommendation \cite{zhang2014lore,bao2012location}, the spread of diseases \cite{liu2015multi}, and contact tracing for hot spots of infectious diseases \cite{126}. 

Data sparsity can become a major challenge when predicting full timelines using publically available data, and can take two forms.
First, the number of users with enough data (full-timelines) to train the model can be low. 
Second, increasing the number of users results in inclusion of users with extremely sparse location timelines. Therefore, this problem of inference of complete location timelines is inherently different from mobility prediction, which must prioritize accuracy of the prediction for the next location. Accordingly, mobility prediction models are often built on data sources such as travel surveys, Call Data Records (CDRs) and Global Positioning System (GPS) trackers which are high-resolution (provide location every few seconds or minutes). However, it is not realistic to have such data in a broad array of contexts; the cost of collecting such datasets, limited attributes associated with individual records, and lack of public availability makes them unsuitable for carrying out large-scale studies for a target population where the impact of location over time is to be studied in relation to various secondary issues. Further, in several emerging real-world modeling applications such as infectious disease transmission models, knowing a person's location at such a high temporal frequency is unnecessary. Instead, locations of where they travel to over the course of a day at a lower resolution (such as every few hours), provide the relevant insight \cite{31}. Therefore, we focus on the challenging problem of constructing the entire mobility timeline of an individual at equal intervals of time from the geo-location associated with social media data, which is generated at-will and therefore can be very sparse.

The main challenge here is that the data used is truly sparse in relation to entire timelines; for example, in six months of social media sourced from the Twitter Application Programming Interface, only 5.4\% (which is what we use in this study) have a Tweet with linked-location at each of the daytime hours in a day (independent of day of the week, over all weeks in the six months). Further, we do not assume any specific information such as from the text/content of posts, or network of users is available. 
To overcome these challenges, we use several known heuristics about location visitation patterns of individuals. We also combine patterns both from an individuals' history, as well as leverage the patterns of similar community members \cite{jurgens2015geolocation,104,139}. Further, we relax criteria about day and week-specificity of location patterns which enables us to use and predict timelines from thousands of social media users with such sparse data -- our method does not require rich training data to learn complex patterns of mobility, and works for a realistic number of users. Finally, our work does not assume any additional information about the demographics, social networks or content of tweets of users, allowing for the adoption in situations where such additional data sources are not available. In accordance with the sparse nature of the data, these combined approaches enable us to infer multiple consecutive missing locations from a user's timeline, and construct a continuous location timeline for individuals using only sparse geo-tags from their Tweets. 

We compare the performance of our model with several models which, although optimized for next-location prediction, are state-of-the-art, and show that intermediate location computing (ILC) has increased accuracy for inferring entire timelines from sparse data. In particular, while deep learning models have good predictive accuracy, we investigate the tradeoff in performance based on amount of sparsity, both in terms of number of users or amount of data per user. By using readily available data to estimate a full mobility timeline at relevant resolutions, this work opens many new opportunities to understand and predict human movement for many domain areas. To the best of our knowledge, this work is the first to use sparse social media data to infer full individual-level location timelines. The specific contributions of this work are:
\begin{itemize}
    \item Developing a framework for filling in entire location timelines at reasonable time steps, with personalized forward and backward timeline prediction.
    \item Prediction of the timeline from truly sparse, but freely available and easily accessible data; with smart use of community data to improve timeline prediction when applicable.
    \item First use of deep learning for inferring timelines from sparse data and assessment of amount of data needed for a deep learning approach to surpass other models.
\end{itemize}

\section{Related Work}
Here we summarize related work in two main categories to clarify differences in data and methodological approaches in other work.
\paragraph{Geo-location data types and sparsity.} As the goal of predicting next location is different than the goal here, the types of data used in such studies include smartphone data including GPS tracking, Wi-Fi, Bluetooth and phone usage \cite{68}, partial GPS tracks from automobiles \cite{122}, and Foursquare check-in data \cite{67}. These models are generally designed for temporally rich data-sets and thus assume that the training data is abundant and collected at frequent time intervals. Even in the case of Foursquare data, though it can be sparse, only dense sequences of data (minimum sequence length of 5 locations) have been used in predictive efforts \cite{feng2018deepmove}. Hence, studies have been concerned with data collected at such densities, or small time intervals (e.g. every 1, 15 or 30 minutes), and individual records below a threshold number of data points are discarded from the study entirely \cite{56,67,70,74,feng2018deepmove}. While this restriction increases confidence in the stay duration of individuals at a location, this typically (appropriately) limits the problem of prediction to only a single missing location in the future. Given the inconsistency of intervals between location tags in an individual's social media timeline and lack of stay duration information, such models can not directly be applied in the context of sparse social media data \cite{tasse2017state}. A method to capture daily habits of individuals using sparse data has been proposed in \cite{137} (varying the amount of phone GPS data ``seen'' by the algorithm). However, the method initially requires training on users with abundant data histories and hence cannot be replicated with data sources such as from Twitter, where both the training and testing datasets are sparse.

Broadly, related social media efforts have been focused on predicting the location of a given social media post, and not missing locations from a timeline \cite{jurgens2015geolocation}. Such studies have also included users with sufficient data and with certain assumptions (e.g. only on those Twitter users who both themselves and their friends are extremely active on Twitter, with at least 100 geo-tagged Tweets in 1 month and assumes that once a user Tweets from a location, they remain at that location until they Tweet again) \cite{104}. Other research which use social media in the domain of mobility focus on Point-of-Interest (POI) and location recommendation, and provide the insight that similar user behavior can be useful \cite{139}. This method uses behavior of similar individuals and distances between pairs of locations to predict the next POI location for an individual. Thus we incorporate this feature of user similarity into the ILC approach to address sparsity issues, and also compare our method to the proposed method for full timeline inference.

\paragraph{Mobility sequence prediction methods.}
There are many model and pattern based methods that have been used to infer movement of individuals. While the focus of these methods has mainly been to predict the next sequence of locations, and cannot be directly compared to our goal of filling in an entire timeline, they have still provided important knowledge about human mobility that can be used in the timeline problem.

Several variations of Markov models, LZ predictors and prediction by partial matching (PPM), as well as a non-linear spatio-temporal prediction framework, have been investigated \cite{118,74,134,asahara2012mixed,136}. These methods focus on modeling the probability of visitation to a future location by probability or frequency of past visits and popular sequences in existing trajectories, each evaluating it's performance on prediction of a next location. Although that is not our goal, we can still make use of such probabilities in our data by incorporating components of the basic Markov model into the ILC model, though in a manner that promotes filling in all missing data, not just next location prediction. Besides, we also explicitly assess performance of each of these these methods on sparse social media data in comparison to our proposed approach where possible, including NextPlace non-linear predictor \cite{136}, Markov Order-0 and Markov Order-1 models \cite{134}. 

More recently, recursive neural networks (RNNs) have been used to predict individual level mobility timelines \cite{feng2018deepmove,yang2017neural,liu2016predicting}. RNN architectures have been used to predict where a user will check-in next \cite{liu2016predicting} and for next location recommendation \cite{yang2017neural}. Another RNN architecture to predict next location in the timeline of an individual has been proposed in \cite{feng2018deepmove}. The model is again focused on predicting next location more accurately, and incorporates modules in the architecture in order to capture more complex multi-scale patterns. As well, despite the fact that this work aims to predict location value in a user's timeline when data is sparse, the work only focuses on predicting the locations in the subset of timelines of users where richer data is available (described in the previous section), and does not address the challenge of inferring the complete timeline of a user. While these new methods provide a fresh approach to addressing the problem of mobility prediction by allowing the model to learn different behaviors on its own as opposed to previous methods where the behaviors of individuals were manually specified, they are not specifically tailored for predicting complete timelines.
However, given the potential for high performance of deep learning models, we do assess what the tradeoff would be, for performance on our task, in terms of data availability (e.g. with what amount of data would a standard deep model perform better than a model incorporating known movement heuristics a priori).

\begin{figure}[t]
\centering
\includegraphics[width=\columnwidth]{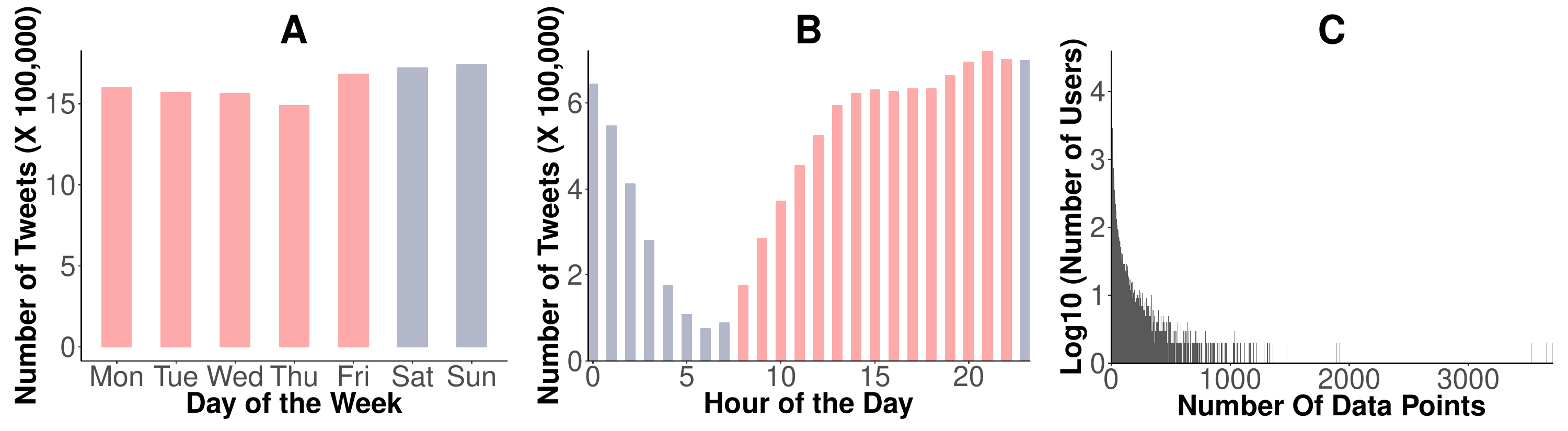}
\caption{A: Number of included Tweets by day of the week (pink: weekdays, blue: weekends). B: Number of Tweets by hour of day (pink: daytime, blue: nighttime). C: Frequency distribution of total data points per user before filling in the timeline. All graphs include all data from all 3 cities.}
\label{data_breakdown}
\end{figure}

\begin{figure}[]
\centering
\includegraphics[width=\columnwidth]{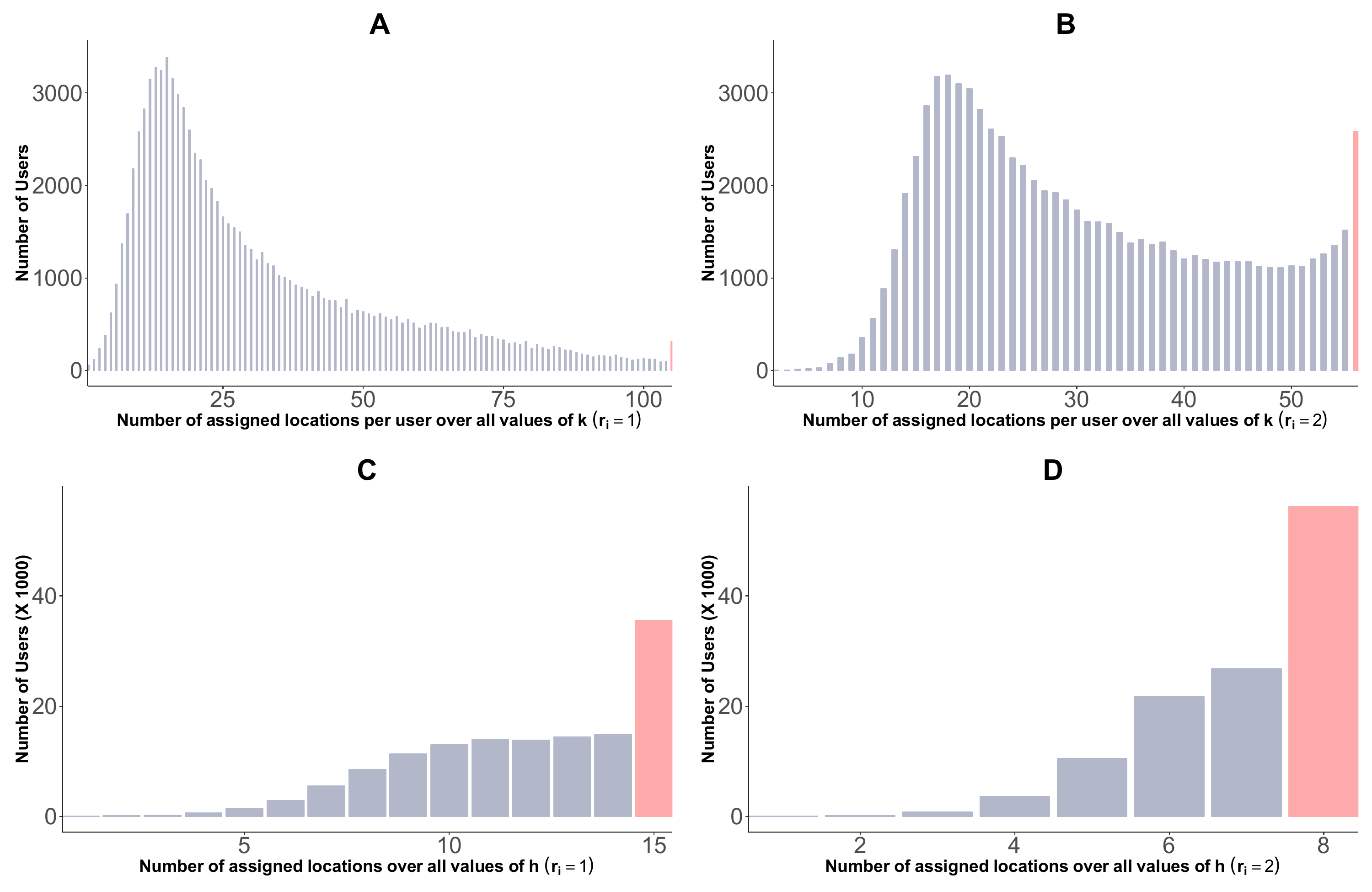}
\caption{A\&B: Frequency distribution of number of assigned locations over all values of $k$ (closest hour of week) for all posts per user (A: resolution $r_i$ = 1 hour, B: $r_i$ = 2). C\&D: Frequency distribution of number of assigned locations per user over all values of $h$ (closest hour of day) (C: $r_i$ = 1, D: $r_i$ = 2). All graphs include data from all 3 cities (excluding users who had no tweet in daytime hours). Users fulfilling inclusion criteria are highlighted in pink.}
\label{data_distrib_k}
\end{figure}

\section{Methods}
\subsection{Datasets}
In order to obtain enough data for training and testing, we used 6 months of publicly available geo-located data from the Twitter API (1st January -- 30th June 2014) for the cities of New York, Washington, DC and San Francisco. We collected all the Tweets containing a `point' geo-location within defined bounding boxes for all three cities. 
The resulting data set consisted of 18,164,503 Tweets by 443,945 users from New York City, 3,385,308 Tweets by 125,873 users from Washington, DC and 1,817,411 Tweets by 111,441 users from San Francisco.
\subsection{Filtering and Preprocessing}
\subsubsection{Spoofed locations}
We identified and excluded any Twitter accounts that represented impossible movements based on Tweet locations and times. A threshold speed of 0.5 miles/minute was used to filter out such Tweets, based on previous work outlining realistic movement patterns \cite{79}, and all accounts with more than 5\% of their Tweets violating the above criteria were excluded. 
A total of 16,582, 3,342 and 2,750 accounts (from each city, respectively) who were removed due to having more than 5\% of their Tweets marked as coming from a spoofed location.
\subsubsection{Grids}
We assessed three grid sizes; 1 $\times$ 1, 0.5 $\times$ 0.5 and 0.1 $\times$ 0.1 miles. For each, we assigned every geo-located Tweet in the dataset to a grid. These grid sizes are based on previous research which has identified perception of how large a neighborhood boundary is for temporary movements such as walking (1 mile) \cite{85,127}. 
Multiple grid sizes were added to assess the impact of grid size on the performance of the method.
A total of 841, 143 and 736 grids(grid size= 1 $\times$ 1 miles), 3,364, 572 and 2,944 grids (grid size= 0.5 $\times$ 0.5 miles) and,  84,100, 14,300 and 73,600 grids (grid size= 0.1 $\times$ 0.1 miles) were created for NYC, DC and SF respectively.
\subsubsection{Temporal Sampling}
Tweet timestamps were adjusted for time zone and daylight savings. 
Included Tweets were distributed across all days of the week evenly (Fig. \ref{data_breakdown}A). For each individual present in the dataset we created separate timelines at resolutions of $r_i$ = 1 and 2 hours. Given the time stamp $t$ of a Tweet, and value $r_i$, $k_w := (h,d)_w$, is computed where $k$ is the closest sampled hour at an interval of $i$ hours from the start of the week, $h$ is the closest sampled hour at an interval of $r_i$ from the start of the day, $d$ is the day of the week and $w$ is the week number since the start of the data. For example, for $r_i=1$, the time of a Tweet made at $t=19:05$ on Tuesday would be assigned $k=43$ i.e 24+19 and $(h,d)=(19,2)$ (assuming week starts on Monday). For $r_i=2$, for the same Tweet, $k=22$ i.e. $12+10$ and $(h,d)=(10,2)$. 
\subsubsection{Stay Duration}
To estimate the stay duration, we interpolated data points from users who made consecutive tweets from the same location within a 6-hour or shorter time period. The maximum value of 6 hours for the interpolation was a conservative estimate chosen based on research showing how long people generally remain in their most visited locations, and that an individual generally spends most of their time in most visited locations \cite{141,125,142}.   
\subsubsection{Home Location}
Individuals are more likely to stay at their home location for longer periods and individuals generally don't change locations at night time \cite{hossain2016precise}.
Consistent with previous studies and Fig. \ref{data_breakdown}B, we consider a location $x$ as the home location of an individual on a day of week $d$, if the individual most frequently tweets from location $x$ between 10 pm of the day of the week $d$ and 8 am of the day of the week $d$+1. Given the sparse nature of the data, for days of the week wherein an individual had no Tweets between 10 pm and 8 am, the home location was assigned where they most frequently Tweeted from between 10 pm and 8 am, irrespective of day of the week. We refer to points in an individual's timeline with location information, either originally from a user or interpolated from a home or stay duration, as \emph{assigned locations}.
\subsubsection{Personal vs Non-personal accounts}
A Twitter account, e.g. @SearchAmerican, that belongs to an organization, as opposed to an individual, is likely to be used by multiple individuals in the organization and hence does not represent the movement patterns of a single individual. To examine the distribution of personal vs. non-personal accounts in our dataset, we used Amazon Mechanical Turk (AMT) labelling on 7,000 randomly selected accounts. Each account label was manually annotated twice by AMT workers as either personal or non-personal accounts. Accounts with conflicting labels were annotated a third time through AMT and the maximum vote used. 98\% of the 7000 randomly selected accounts were identified as personal accounts. Cohen's kappa score of the annotators was 93.0\% \cite{78}. Given the overwhelming majority of accounts were identified as personal, it was assumed that most non-personal accounts must have been removed during the `spoofed location' filtering stage. The 2\% non-personal accounts from the 7,000 set were removed from the study but it was deemed unnecessary to label the remaining accounts. After this stage, no information (e.g. the Twitter handle) which could link back to an individual Twitter account holder was retained. 

\subsubsection{Description of Included Users}
We define a relaxed inclusion criteria to ensure that the performance of all methods is being tested on users with sparse timelines. From here onward we define two notations: $x_{k_{w}}(u)$ is the assigned location for a user $u$, at time $k_w$, with $k_w$ as above.  $k_w$ can also be interchangeably written as $(h,d)_w$, with $h$, $d$ and $w$ as above, or as $q$, which represents the index of $k_w$ in the sampled timeline. Inclusion criteria were defined as follows: given the timeline of a user, the user must have at least 1 assigned location for each $h$ during daytime hours (8am-10pm;non-nighttime hours as defined in the section "Home Location"), irrespective of $d$ and $w$. This means that at a resolution of $r_i$ = 1 hour, all users were included in our analysis who, after interpolation of stay duration had at least 15 assigned location data points (8am-10pm) in the entire duration of the dataset over all distinct $h$. For $r_i$ = 2, the number was 8 assigned location data points.
This resulted in 29,491, 4,947 and 1,119 users ($r_i$ = 1) and 45,710, 8,083 and 2,395 users ($r_i$ =2) from New York, Washington, DC and San Francisco respectively.
Defining a relaxed inclusion criteria based on distinct $h$ instead of distinct $k$ enabled us to include orders of magnitude more users (Fig. \ref{data_distrib_k}) and allowed us to include up to 45\% user (in NYC) who made a tweet during daytime hours. The above selected users had on average 82.8\% ($r_i$ = 1) and 72.0\% ($r_i$ = 2) of their daytime timelines with no assigned location.

\subsection{Individual Timelines}
In this section, we first discuss prediction of a missing location in a user's timeline at time $k$, if location information of the user is available at both $k+1$ and $k-1$.
As described in earlier work, the movement of individuals is not entirely random and certain features can be extracted to predict an individual's location based on his past behavior \cite{57}. Moreover, people often move in groups, and individuals with similar interests follow similar movement patterns \cite{81}. Accordingly, here we model the behavior of individuals as a combination of: i) personal behavior represented as (subscript $I$), and ii) community behavior (subscript $C$). Personal behavior is further modeled using three behaviors: i) Next Location subscript $(I,a)$, ii) Previous Location subscript $(I,b)$, and iii) Independent Location subscript $(I,c)$. Each of these three behaviors are further treated as either i) day of the week and hour of the day specific (superscript $WS$), ii) workday (weekday) or non-workday (weekend) and hour of the day specific (superscript $RS$), and iii) only hour of the day specific (superscript $HS$). These three stratifications were created because of the extremely sparse nature of the dataset in which we rarely observe users who have at least 1 location value present for all days of the week and hours of the day. 
For the following section we define: any $P$ represents a list of locations and their corresponding probabilities for a given user at a given time. $P(x_{k_w}==j)$ thus represents the probability corresponding to location $j$ at time $k_w$, and $P(x_{k_w})$ represents the list of all possible locations and their corresponding probabilities at time $k_{w}$. 

\subsubsection{Next Location} 
%
%
Given the location $x_{{(k+1)}_w}(u)$ is missing, and given the location $x_{k_w}(u)=i$, we calculate the conditional probabilities of all the possible locations of a user $u$ at time $(k+1)_w$. This probability is calculated by taking into account that people often follow specific patterns of mobility. For example, in the evening at 7pm, given that an individual is at a grocery store, the next location of an individual will likely be his home. Given that the same individual is at home at 7pm, the individual could either choose to stay home or to go out (e.g. to a restaurant or bar). Given that the time period is assumed to be 1 week, as contended in previous work, these conditional probabilities are specific for each sampling time on a given day and day of the week, irrespective of the date \cite{56}. 
Then, for all possible locations $j$, $P_{(I,a)}^{WS} (x_{(k+1)_{w'}}=j)$ of a user, given $x_{k_{w'}}=i$, is defined as: 
\begin{small}
\[
\frac{\sum\limits_{w} (x_{(k+1)_w}==j | x_{k_w}==i)} {\sum\limits_{w} (x_{k_w}==i)}
\]
\end{small}
For ${RS}$ we calculate similar proportions, but relax the conditions by additionally accounting for days which are of the same type, i.e workday or non-workday, when calculating the proportions. $P_{(I,a)}^{RS} (x_{(h+1,d')_{w'}}=j)$ is thus defined as:
\begin{small}
\[
 \frac{\sum\limits_{w} \sum\limits_{d=DT(d')} (x_{(h+1,d)_w}==j | x_{(h,d)_w}==i) } {\sum\limits_{w} \sum\limits_{d=DT(d')} (x_{(h,d)_w}==i)}
\]
\end{small}
where $DT(d')$ returns the list of type of days i.e weekdays or weekends, as $d'$. $P_{(I,a)}^{HS}$ completely removes the condition of the proportion being specific to the day of the week (instead of $d=DT(d')$, we consider all $d$).

\subsubsection{Previous Location} 
As a reciprocal of Next Location prediction wherein we used $x_{k_w}$ to predict $x_{(k+1)_w}$, here we predict $x_{k_w}$ conditioning over the location value at $x_{(k+1)_w}$. 
$P_{(I,b)}^{RS}$ and $P_{(I,b)}^{HS}$ are calculated similarly using relaxed conditions of day of the week, as defined for $P_{(I,a)}^{RS}$ and $P_{(I,a)}^{HS}$.

\subsubsection{Independent Location}
Several locations which an individual visit are specific to the day and time regardless of where the individual is coming from or where they plan to go next. For example, for a weekly meeting or a class at 11am on Tuesday, an individual will be in the location of the meeting or the class irrespective of his previous or next location. To incorporate these patterns, we calculate probabilities for ``Independent Location'': the probability of a user being in any location $j$ at time $k := (h,d)$, $P_{(I,c)}^{WS}$. This is defined as the proportion of times the user was at location $j$ at time $k$, in the dataset. $P_{(I,c)}^{RS}$ of a user being in any location $j$ at time $k := (h,d)$ is defined as the proportion of times the user was at location $j$ during hour $h$ and days of the week similar to $d$ i.e (weekday or weekend). And, $P_{(I,c)}^{HS}$ is defined as the proportion of times the user was at location $j$ during hour $h$ in the dataset.

Combining lists of all probabilities in the individual's ($WS$) behavior gives: 
\begin{small}
\[
P_I^{WS}=(\lambda_a*P_{(I,a)}^{WS}+\lambda_b*P_{(I,b)}^{WS}+P_{(I,c)}^{WS})/3
\]
\end{small}
where $\lambda_a$ and $\lambda_b$ are information loss factors defined later in the Intermediate Location Computing section. 
Probabilities of visit to each location, from all behaviors, are summed to generate a single list of locations and their corresponding probabilities:
\begin{small}
\[
P_I= (P_I^{WS}+P_I^{RS}+P_I^{HS}) / 3
\]
\end{small}

\subsection{Community Behavior}
Individuals with similar interests, or those working or living in the same demographic have a higher chance of visiting similar locations \cite{56}. Hence, we maximize the use of the data by also including information about individuals who have shown to follow similar mobility patterns. For each individual, we identify individuals who have similar mobility patterns, via a similarity factor. This factor, $s(u1,u2)$, is defined as the probability that another individual $u2$ will be in the same location as the individual under consideration $u1$ at any given time: 
\begin{small}
\[ 
\frac{\sum_w \sum_k (x_{k_w} (u1)==x_{k_w} (u2))} { \sum_w \sum_k (!NULL(x_{k_w} (u1)) \& !NULL(x_{k_w} (u2)))}
\]
\end{small}
Using the similarity factor defined above, we calculated community behavior (probability list for locations at a time $k$) using the top $m$ users in the dataset with the highest similarity factor for a given individual via:
\begin{small}
\[
P_C(x_{k_w}=j)= \sum_{u=1}^m s(u)*(x_{k_w}(u)==j)
\]
\end{small}
Combining individual and community behavior then gives: 
\begin{small}
\[
P(x_{k_w}=j)= (1-\beta_k)*P_I(x_{k_w}=j) + \beta_k*P_C(x_{k_w}=j) 
\]
\end{small}
where $\beta_k$ defines the hour and day of week specific effect of community behavior on an individual. To account for varying behavior of an individual during a week, we generated separate lists of similar users for weekdays and weekends. We also examined a range of values for $m$ (0, 1, 2, 5, 10, 20, 50), to identify the minimum number of similar users for maximizing prediction accuracy. 

\subsection{Intermediate Location Computing}
Given the sparse nature of social media, in most instances there are multiple consecutive missing location data points in an individual's timeline. Thus the issue of predicting location at $k_w$ if either or both $(k+1)_w$ and $(k-1)_w$ are missing will arise. Hence we introduce the concept of Intermediate Location Computing. For simplicity, we will only define the procedure to identify the intermediate location at sampled time $(k-1)_w$ (which is used to calculate of $P_{(I,a)}^{WS}$). A similar approach can be used to identify the location at time $(k+1)_w$ (which is used to calculate $P_{(I,b)}^{WS}$).


Broadly, our problem is that a location exists at time $(k-n)_w$ such that no location data for an individual is present between $k_w$ and $(k-n)_w$. To address this, we use location data at $(k-n)_w$ to iteratively predict intermediate locations of the individual at times $(k-n+1)_w$ until we reach $(k-1)_w$. We define the function $Inter(L1,L2)$, which for a specific time point, takes in two lists $L1$ and $L2$, and returns the location which has the maximum probability in list $L1$, and if no location exists, returns the location with maximum probability in list $L2$. Here $L1$ is $P_{(I,a)}$ and $L2$ is $P_{(I,c)}$. In simple terms, at each step, we first identify the most probable location using Next Location. If no location data exists, we resort to identifying the most probable location using Independent Location.  




Given that locations at $(k+1)_w$ and $(k-1)_w$ predicted using this method are only probable locations and successive predictions will decrease certainty, we multiply by an information loss factor $\lambda$ to account for loss in information in calculating intermediate locations. This factor $\lambda$ is defined as: $\lambda=(1-\alpha)^{(n-1)}$, where $n$ is the number of steps required to reach the nearest available point with an available location, and $\alpha$ is a constant information loss on each step. 

\begin{figure}[t]
\centering
\includegraphics[width=\columnwidth]{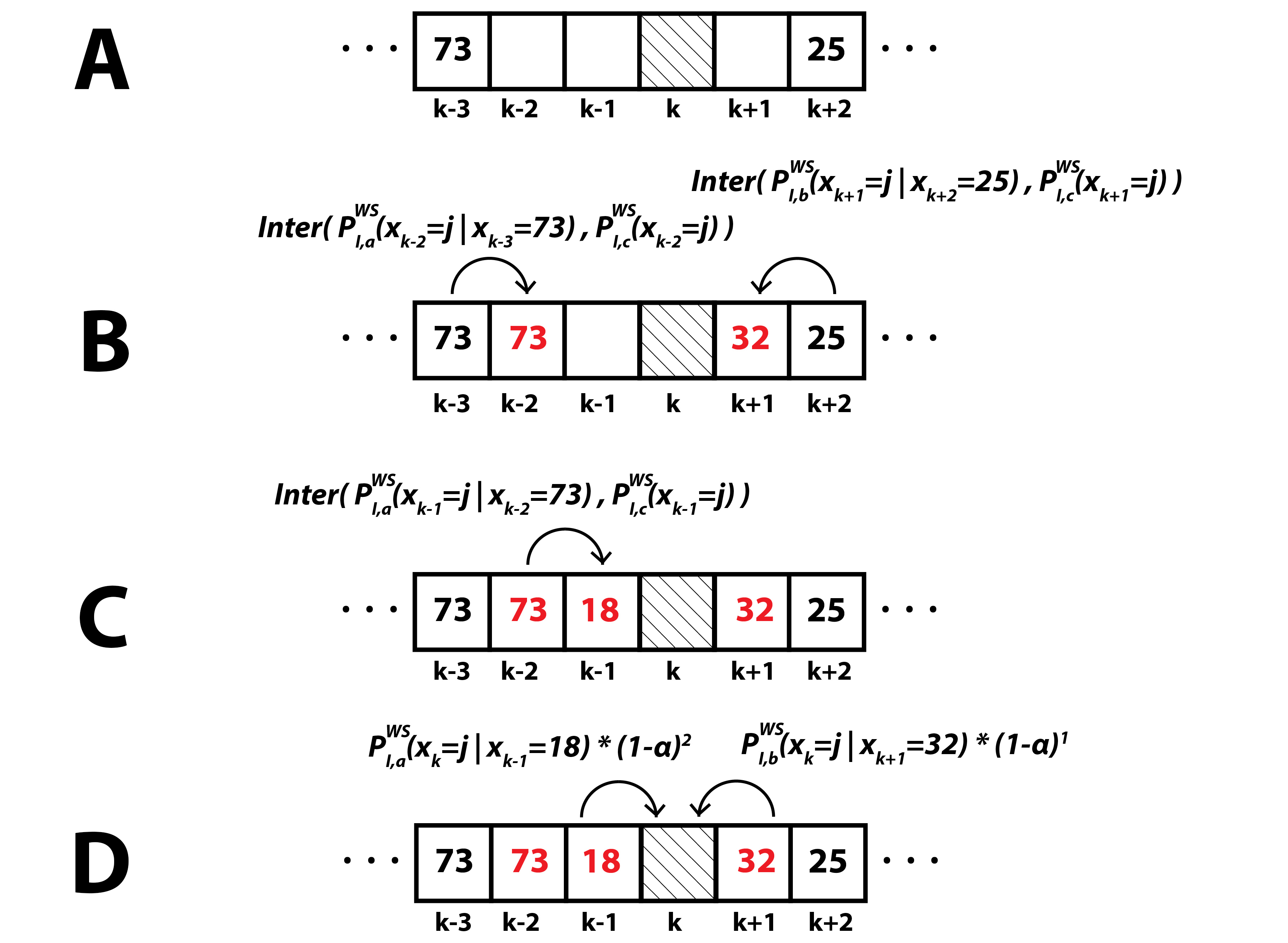}
\caption{Intermediate Location Computing algorithm illustration. A: Timeline of an individual for a week $w$, between $k-3$ and $k+2$. Location data for the individual is missing for $k-2$ to $k+1$. Shaded area shows the location to be predicted. B: Intermediate locations (red) calculated after first iteration. C: Intermediate locations after second iteration. D: Effect of information loss on $P_{(I,a)}^{WS}$ and $P_{(I,b)}^{WS}$.}
\label{ILC_fig}
\end{figure}

This approach to identify loss in information in sequential predictors has been used in the past, particularly in dynamic belief models \cite{140}. The basic idea is that at each sequential prediction there is a probability of $\alpha$ that the prediction will be incorrect. Iterating this for a data point present $n$ steps away makes the overall probability of correct prediction $(1-\alpha)^{(n-1)}$. In the example given in Fig. \ref{ILC_fig}, when finally calculating the location at $k_w$, given that the value of $n$ for the left side is 3, $P_{(I,a)}^{WS}$ is multiplied by $(1-\alpha)^{(3-1)}$. Similarly given that $n$ for the right side is 2, $P_{(I,b)}^{WS}$ is multiplied by $(1-\alpha)^{(2-1)}$. The example in Fig. \ref{ILC_fig} demonstrates the steps performed to compute the intermediate locations for ($WS$). We use the same method to calculate intermediate locations for ($RS$) and ($HS$) probabilities.

The complete method to construct complete mobility timeline of a given user is summarized algorithm \ref{algo}. In the algorithm, as defined above, we replace $k_w$ with $q$, to represent the index of each time step in the timeline. Further given a timeline $T$ of an individual, the location $x_q$ is the $q$ element in $T$, i.e $T[q]$. 

\begin{algorithm}
\caption{Constructing complete mobility timeline using ILC}
\label{algo}
\begin{flushleft}
\textbf{Input}: Timeline $T$ of user $u$, community behavior, $P_C$, of similar users at each time step, effect of community behavior $\beta$ and $\lambda_a$ and $\lambda_b$ for each $q$\\
\textbf{Output}: Complete timeline $T_{complete}$
\end{flushleft}
\begin{algorithmic}[1]

\State $T_{complete} \leftarrow T$
\\
\For {each behaviour $S$ in $[WS,RS,HS]$}
    \State $T_a^S \leftarrow T_b^S \leftarrow T$
    \For{ $q$ in $1:length(T_a^S)$ }
        \If {$NULL(T_a^S[q])$}
        \State $T_a^S[q] \leftarrow inter(P_{I,a}^S{(x_{q}|x_{q-1}=T_a^S[q-1])},P_{I,a}^S(x_{q}))$

        \EndIf
    
    \EndFor
    
    \For{ $q$ in $length(T_b^S):1$ }
        \If {$NULL(T_b^S[q])$}
        \State $T_b^S[q] \leftarrow inter(P_{I,b}^S{(x_{q}|x_{q+1}=T_b^S[q+1])},P_{I,a}^S(x_{q}))$

        \EndIf
    
    \EndFor

\EndFor
\\
\For {each $x_q$ in $T$}
    \If {$NULL(x_q)$}
        \For {each behaviour $S$ in $[WS,RS,HS]$}
            \If{$NULL(x_{q-1})$}
                \State $P_{I,a}^S(x_q) \leftarrow P_{I,a}^S{(x_{q}|x_{q-1}=T_a^S[q-1])}$
            \Else
                \State $P_{I,a}^S(x_q) \leftarrow P_{I,a}^S{(x_{q}|x_{q-1}=T[q-1])}$
            \EndIf
            
            \If{$NULL(x_{q+1})$}
                \State $P_{I,b}^S(x_q) \leftarrow P_{I,b}^S{(x_{q}|x_{q+1}=T_b^S[q+1])}$
            \Else
                \State $P_{I,b}^S(x_q) \leftarrow P_{I,b}^S{(x_{q}|x_{q+1}=T[q+1])}$
            \EndIf
            
            \State $P_I^{S}(x_q)=(\lambda_a*P_{(I,a)}^{S}(x_q)+\lambda_b*P_{(I,b)}^{S}(x_q)+P_{(I,c)}^{S}(x_q))/3$
        \EndFor
        \State $P_I= (P_I^{WS}(x_q)+P_I^{RS}(x_q)+P_I^{HS}(x_q)) / 3$
        \State $T_{complete}[q] \leftarrow arg.max((1-\beta_q)*P_I(x_q) + \beta_q*P_C(x_q)) $
    \EndIf
\EndFor

Return $T_{complete}$
\end{algorithmic}
\end{algorithm}

\subsection{Training and Testing Data and Optimization}
To select the training data for the entire prediction we, randomly and uniformly across all distinct values of $k$, sampled 70\% of the data from each user. It should be noted that the test set contains the 30\% location data of each user which was not used in calculating the conditional probabilities or training the model. Further, the data spans only the daytime hours wherein an individual is changing location most frequently. The performance of the model was calculated only on this test data as not to bias the performance of the method towards sampled times where an individual is static (nighttime hours). 

Using the training data, we calculated $P_I$ and $P_C$ lists for every individual, at each time resolution. These probabilities are then used to optimize the value of $\beta_k$ and $\alpha$. For simplicity, we optimize a fixed value independent of a user or a sampling time for $\alpha$, but $\beta_k$ is user, day of the week and hour of the day specific as we would expect the contributions of community behavior to vary at different times and for different people. To select the optimal values of $\beta_k$, 
we vary $\beta_k$ from 0 and 1 ( intervals of 0.05) and select the $\beta_k$, for a given $k$, that maximizes prediction accuracy on the training data.
$\alpha$ was optimized in a similar way, but only using $P_I$ (inclusion of $P_C$ would have resulted in concurrent optimization of both $\alpha$ and $\beta_k$). The value of $\alpha$ as 0.1 performed well on the training set, and was used in study.

\subsection{Evaluation Versus Baseline Models}
For fair comparison and to ensure that the variation in performance is only due to the inference power of the models and not due to variation in training data, all baseline models were trained using the same training data for each user (post processed form of data) as used for ILC, and the performance of the models was tested on the same test set. 

\subsubsection{Home-Work location Model}
It has been shown that periodic behavior accounts for up to 70\% of an individual's movement \cite{133}. Given that the periodic behavior 
Hence, the first baseline model 
assumes users follow a simple periodic behavior, switching between two locations: their inferred home and work locations. Using the training dataset, we computed and assigned a single home (nighttime) and a work (daytime) location for each individual by identifying the most frequent location a user is present in
between 10pm and 8am, and between 8am and 10pm.

\subsubsection{Markov Models}
Markov models have been widely used to predict individual level mobility patterns \cite{112,134}. An Order-0 Markov model
identifies the most frequent location a user is in during a given hour of the day, regardless of where the user came from or is going\cite{134}.
The Order-1 Markov model, 
given the location $x$ of an individual at time $k$, 
identifies the most frequent location the individual visits at time $k+1$ if they were at $x$ during time $k$. Due to sparsity of data, multiple missing locations are predicted iteratively.
i.e. each subsequent location at $k+n$ is predicted using the previously predicted location at $k+n-1$.
For fair comparison, we use a fall-back version for both Markov models which first computes the ($WS$) likelihoods. If no location data exists, the model falls back to ($RS$) likelihood, and then to ($HS$) likelihood. 


\subsubsection{Collaborative Point-of-Interest Recommendation Model}
The Point of Interest (POI) recommendation model was initially presented in \cite{139}, to recommend locations of interests of individuals using data from Location Based Social Networks (LBSNs). 
The model, in addition to using geographical distance between locations, first identifies close users both based on the social network (friends/followers) of an individual as well as those who follow similar movement patterns, and uses their location to predict the individuals location. 
In line with the conclusion of the original work, that social ties are not strong predictors, and given that we are not assuming that the location data for the social network of individuals is available, we model the movement of an individual using the geographical distances between locations and location data of users who follow similar movement patterns.  
Geographical influence is modelled based on a power-law distribution between successive data points, while location of similar users is calculated similar to the community behavior part of our method. 

\subsubsection{NextPlace: Spatio-Temporal Non-linear Model}
This spatio-temporal non-linear ``NextPlace'' prediction model uses a non-linear framework for predictions and unlike Markov models, which predict the next location at time $k+1$ using historical movement patterns, or the community based methods, which use location data of similar users, 
uses the history of trips to the same location to predict when an individual will be in the same location the next time. \cite{136} 
The method first identifies the start time and stay duration of each trip, then embeds the timeseries in a multidimensional space by adding multiple instances of the timeseries with delays to account for non-linearity. Then, the start times and stay durations of the user's next visit are averaged to predict when and for how long the next visit to the location will happen. In our implementation, we used the delay as the smallest temporal unit in our study (i.e 1 and 2 hours for $r_i$=1 and 2). Given the sparsity of data, we define the start time when an individual makes a tweet from a location, and stay duration is either inferred as described in the preprocessing section of paper, or assumed to be either 1 or 2 hours based the value of $r_i$.


\begin{figure}[t]
\centering
\includegraphics[width=0.7\columnwidth]{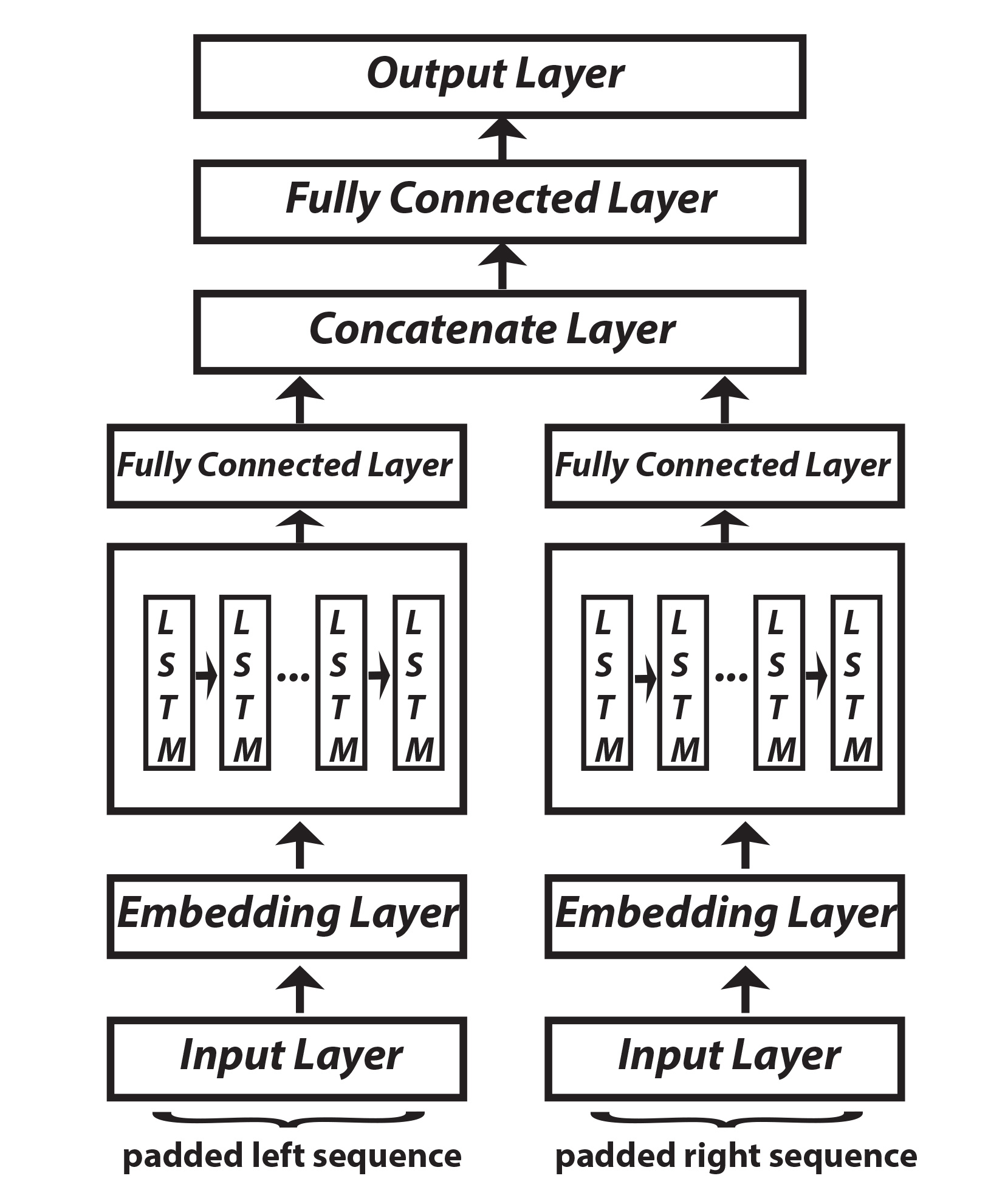}
\caption{Architecture of Recursive Neural Network.}
\label{RNN_fig}
\end{figure}

\subsection{Recursive Neural Network}
RNNs and specifically LSTMs (Long Short-Term Memory blocks) have been gaining popularity due to their strength in identifying and utilizing complex sequences of information to make future predictions. For the domain of mobility prediction, this provides a contrast to other work in which the heuristics for mobility modelling are self specified. Hence, here we also study the utility of an RNN architecture in constructing entire mobility timelines of individuals in the context of sparse location data. Fig. \ref{RNN_fig} shows the architecture of the network. We use a basic architecture, similar to those used in previous mobility and sequence prediction work \cite{feng2018deepmove}, but adapted for full timeline location inference. Specifically, instead of using separate inputs for current and historical trajectories of location, due to the sparse nature of data we input a single trajectory of all locations. Secondly and more importantly, here instead of only using a historical sequence of locations (left padded input), [$x_{(k-n)_w}$,... $x_{(k-2)_w}$, $x_{(k-1)_w}$] to predict $x_{k_w}$, we also use the future sequence of location (right padded input), [$x_{(k+1)_w}$, $x_{(k+2)_w}$,... $x_{(k+n)_w}$], to maximize the utility of sparsely available data and predict a location value for ever missing time step (not just next step). Thus, the architecture comprises of left and right padded input layers which are fed to embedding layers to convert sparse inputs into dense representations. The outputs from the embedding layers are then input to recurrent units comprising of an array of LSTM units. The LSTM outputs are then passed through a fully connected layer and concatenated before being passed through a fully connected layer to interpret the output and make prediction. 
All fully connected layers use rectified linear unit activation except for output layer which uses softmax activation. The model uses categorical cross entropy loss function and uses Adam optimizer to update weights in the network. The model is trained using the training dataset. 10\% of the training set is set aside for validation. After each epoch, the performance of the model is tested on the validation dataset. 
The training is stopped when no improvement in prediction accuracy of validation data is observed. Though architectures can be further augmented with other types of modules to model further complexities, the comparison here is meant to evaluate the pure heuristic versus deep learning approaches.

\begin{table*}[t]
\centering
\caption{Overall prediction accuracy (\%) and average percentage of filled timelines(written in \{\}) for baseline models and Top 1 and Top 3 locations predicted by the intermediate location computing model.}
\begin{tabular}{llllllllll}
\toprule
City           & $r_i$   & Top 1 & Top 3 & RNN &  Home-Work & Markov O(0) & Markov O(1) & POI   & NextPlace \\
\midrule
NYC  & $r_i$=1 & 72.69\{100\} & 82.35\{100\} &  73.09\{100\} &   65.54\{100\}     & 64.65\{100\}       & 26.39\{32.70\}       & 15.59\{56.04\} & 0.17\{18.07\}      \\
               & $r_i$=2 & 64.78\{100\} & 77.38\{100\} & 59.33\{100\} & 59.28\{100\}     & 57.98\{100\}       & 32.56\{48.69\}       & 19.11\{76.75\} & 0.21\{28.93\}      \\
DC & $r_i$=1 & 75.08\{100\} & 83.61\{100\} & 74.58\{100\} & 66.91\{100\}     & 65.76\{100\}       & 27.75\{32.29\}       & 31.27\{70.60\} & 0.11\{17.23\}      \\
               & $r_i$=2 & 68.85\{100\} & 79.57\{100\} &63.27\{100\}& 62.35\{100\}     & 60.64\{100\}       & 34.13\{48.79\}       & 34.56\{82.56\} & 0.19\{28.36\}      \\
SF  & $r_i$=1 & 77.20\{100\} & 86.28\{100\} &76.26\{100\}& 67.74\{100\}     & 67.21\{100\}       & 16.78\{30.12\}       & 35.49\{60.24\} & 0.15\{17.57\}      \\
               & $r_i$=2 & 70.78\{100\} & 82.06\{100\} &64.78\{100\}& 63.66\{100\}     & 62.91\{100\}       & 19.52\{43.72\}       & 32.69\{67.69\} & 0.22\{28.50\}\\ 
\bottomrule               
\end{tabular}

\end{table*}

\begin{table}[]
\centering
\caption{Prediction accuracy (\%) for Top 1 (T1) and Top 3 (T3) locations predicted by the ILC model by grid size. ($g$) represents a grid size of $g \times g$ miles.}
\begin{tabular}{llllll}
\toprule
City & $r_i$   & T1(0.5) & T3(0.5) & T1(0.1) & T3(0.1) \\
\midrule
NYC  & $r_i$=1 & 65.64       & 75.71       & 54.23       & 64.07       \\
     & $r_i$=2 & 59.29       & 71.65       & 46.06       & 57.96       \\
DC   & $r_i$=1 & 67.32       & 77.65       & 54.27       & 64.10       \\
     & $r_i$=2 & 60.19       & 72.59       & 46.85       & 58.23       \\
SF   & $r_i$=1 & 70.86       & 80.97       & 57.37       & 67.26       \\
     & $r_i$=2 & 63.37       & 75.47       & 48.07       & 59.81      \\
\bottomrule
\end{tabular}

\end{table}

\begin{figure}[t]
\centering
\includegraphics[width=\columnwidth]{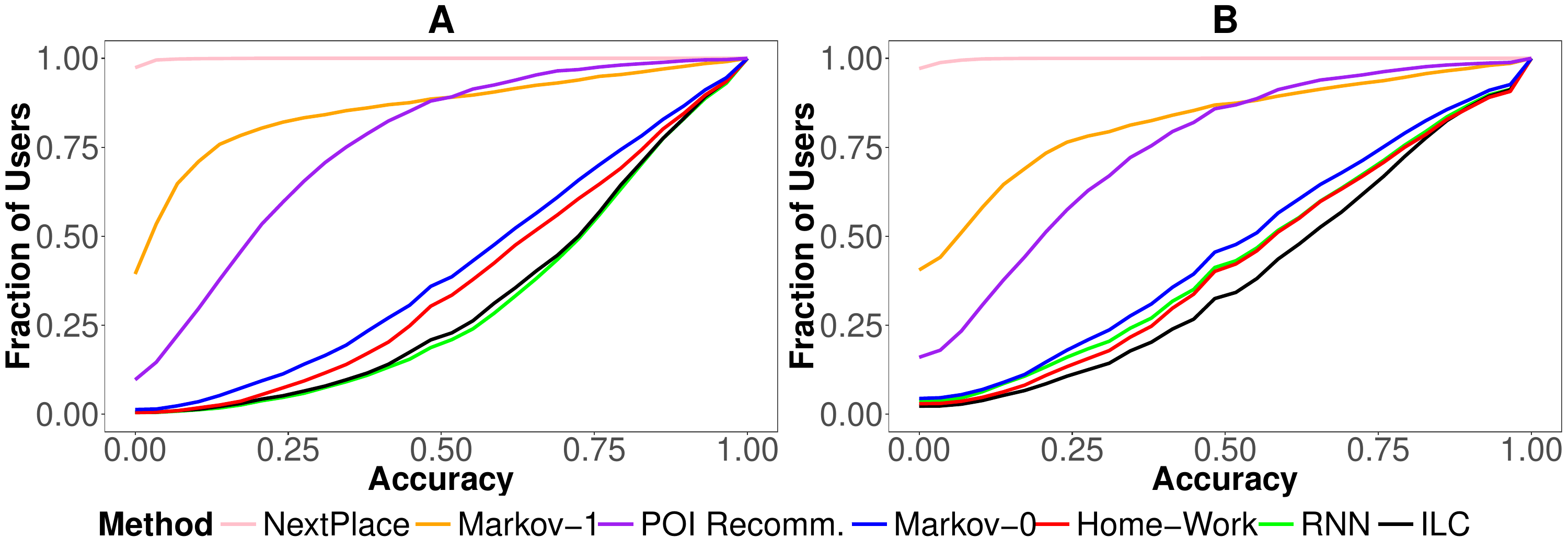}
\caption{Cumulative fraction of users vs. prediction accuracy for ILC and baseline models, $r_i=1$, (A) and 2 (B).}
\label{cdf_accuracy}
\end{figure}

\begin{figure}[t]
\centering
\includegraphics[width=0.5\columnwidth]{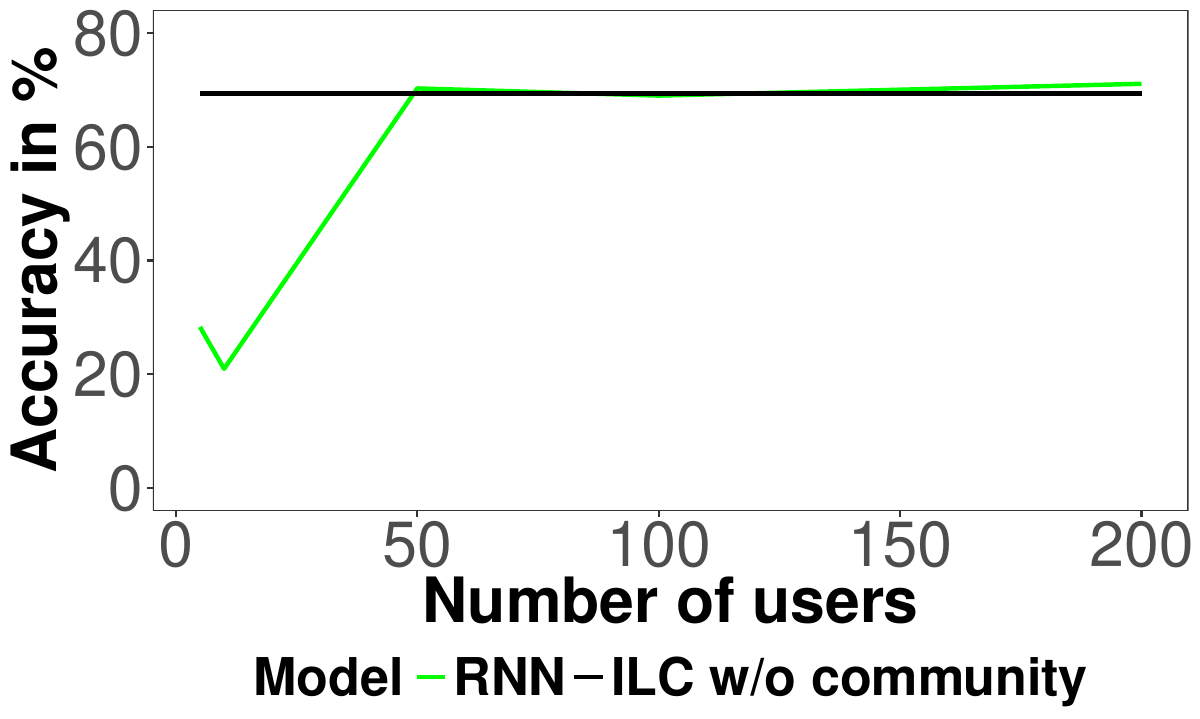}
\caption{Prediction accuracy of ILC (no community data) and RNN with number of users used to train the model for NYC $r_i=1$. Values calculated at \# of users=5,10,50,100,200 using mean of 10 replications. At \char`\~50 users RNN performance comes close to ILC, and by 200 users the RNN model surpasses ILC. Accuracy values for training with all available users are in Table 1.}
\label{rnn_num_user}
\end{figure}

\section{Results}

\subsection{Comparative Performance of Methods}
The ILC, RNN, Home-Work and Markov-0 models predicted a location value for every missing data point in the dataset (Table 1). 
Amongst the remaining methods, the NextPlace algorithm `filled-in' the least number of missing data points. ILC and RNN outperformed all baseline models across all cities (Table 1). 

For $r_i=1$, RNN slightly outperformed ILC in only NYC when considering the overall performance of methods on test data points (Table 1). When analyzing prediction accuracy per user in the test set, RNN slightly outperformed ILC (Fig. \ref{cdf_accuracy}). For $r_i=2$ ILC outperformed RNN across all cities both when considering overall accuracy on test data and accuracy per user. Additionally, for $r_i=2$, despite RNN outperforming Home-Work location model when considering overall prediction accuracy on test data (Table 1), it performed slightly worse than Home-Work location method when considering accuracy per user (Fig. \ref{cdf_accuracy}B). 

Amongst the heuristic-based baseline models, simpler models outperformed more complex models. This was mainly because they were able to predict a location value for larger number of missing data points. The Order-0 Markov and Home-Work location model resulted in similar prediction accuracy and outperformed the remaining baseline models. 
In contrast to previous work, the Order-1 Markov model had a lower prediction accuracy as compared to Order-0 Markov, largely because it was only able to predict a location value for one-third of the data points in timelines at $r_i$=1 and one-half of the data points in timelines at $r_i$=2. The time-dependent POI recommendation model outperformed Order-1 Markov model in SF and DC and underperformed in NYC. This is consistent with the fact that as shown in Fig. \ref{smilar_user}B, SF and DC had higher similarity between locations of individuals as compared to NYC. Additionally, the POI model was able to predict a much larger portion of users' timelines as compared to the Markov-1 model, yet accuracy values for both methods were close. The NextPlace method based on a non-linear spatio-temporal framework had the least predictive power given the fact that it relies largely on stay duration information. Given the lack of this information in social media, the model was scarcely able to predict missing location values.


While the baseline heuristic based methods have been optimized for different data types, in general, ILC specifically addresses the challenge of sparse data by incorporating a wide range of components. The simpler components help predict a location value for each missing point, while the more complex components help identify complex movement behaviours. 


Comparing ILC with RNN shows that RNNs are powerful methods that can out perform traditional heuristic based methods. However, we see that in low data settings, heuristics can be used to outperform the deep learning approach (e.g. when predicting at less frequent time intervals, or when a lower number of users are available to train the model). This is evident in Fig. \ref{rnn_num_user}, where despite RNN outperforming ILC in NYC at $r_i=1$, if trained on a fewer number of users, it under performs. Also in Fig. \ref{cdf_accuracy} we see that the RNN requires data from more users to achieve the same accuracy as ILC, when considering $r_i$=2. However with increases in the amount of training data, the RNN outperforms ILC. This is due to the fact that this implementation of ILC only uses a maximum sequence length of two time steps, RNNs can learn larger and more complex sequences of locations. Additionally RNNs can also learn longer sequences of location data of similar users and help improve prediction. Decision between selecting one over the other is based on the goal of the study and the availability of data. If sparse data for a large number of users is available, then an RNN approach should be preferred. But if the goal of the study is to maximize the number of users for which complete timelines can be constructed by sampling their locations at less frequent time intervals, or if the number of available users is low, then a heuristic based method like ILC should be preferred given that it does not need data to learn patterns. 

\subsection{Effect of Community Behavior}
We found that the effect of community behavior is consistently higher on an individual's mobility patterns during weekends as compared to weekdays ($\beta_k$ higher on weekends across all cities and $r_i$). The average value for $\beta_k$ during weekdays ranged from 0.449 (NYC, $r_i$ =1) to 0.466 (DC, $r_i$=2), while during weekends ranged from 0.456 (NYC, $r_i$ =1) to 0.492 (DC, $r_i$=2).

\begin{figure}[t]
\centering
\includegraphics[width=\columnwidth]{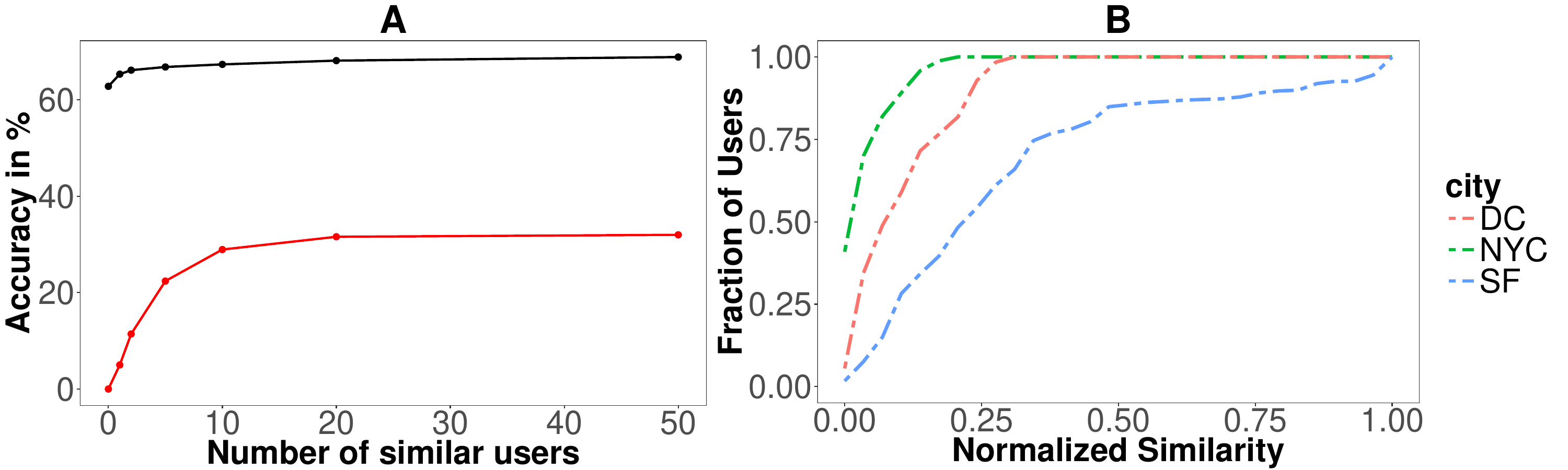}
\caption{A: Prediction accuracy (\%) vs. number of similar users ($m$). Model based on individual and community behavior (black), and community behavior only i.e $\beta=1$ (red). All values were computed for $r_i=2$ and using Washington, DC data. B: Aggregated similarity factors of closest users ($\sum_{u=1}^m{s(u)}$) vs. fraction of users by city. X-axis normalized by dividing by the maximum aggregated similarity factor for a user in the dataset.}
\label{smilar_user}
\end{figure}

We observe that inclusion of community data helps the performance of the method and the main improvement is seen when the first similar user is accounted for (Fig. \ref{smilar_user}A). Moreover, after $m$ = 20, 
accuracy improvements begin to plateau with more $m$ (the inclusion of $m$ closest individuals to compute community behavior will work best for individuals who have high similarity values with other individuals and are not outliers in terms of their mobility) justifying the use of $m$ = 50 in our method.

\subsection{Performance of ILC in Different Settings}
We observe that for ILC performance decreases as the interval ($r_i$) increases from 1 to 2 hours (Table 1), and as the grid size decreases (Table 2), which is inline with the findings of \cite{cuttone2018understanding} that at larger time intervals and smaller grid sizes there is a higher associated uncertainty . Similar trend is observed for RNN as increase in $r_i$ from 1 to 2 hours decreases the overall training data for the model.

\begin{figure}[]
\centering
\includegraphics[width=\columnwidth]{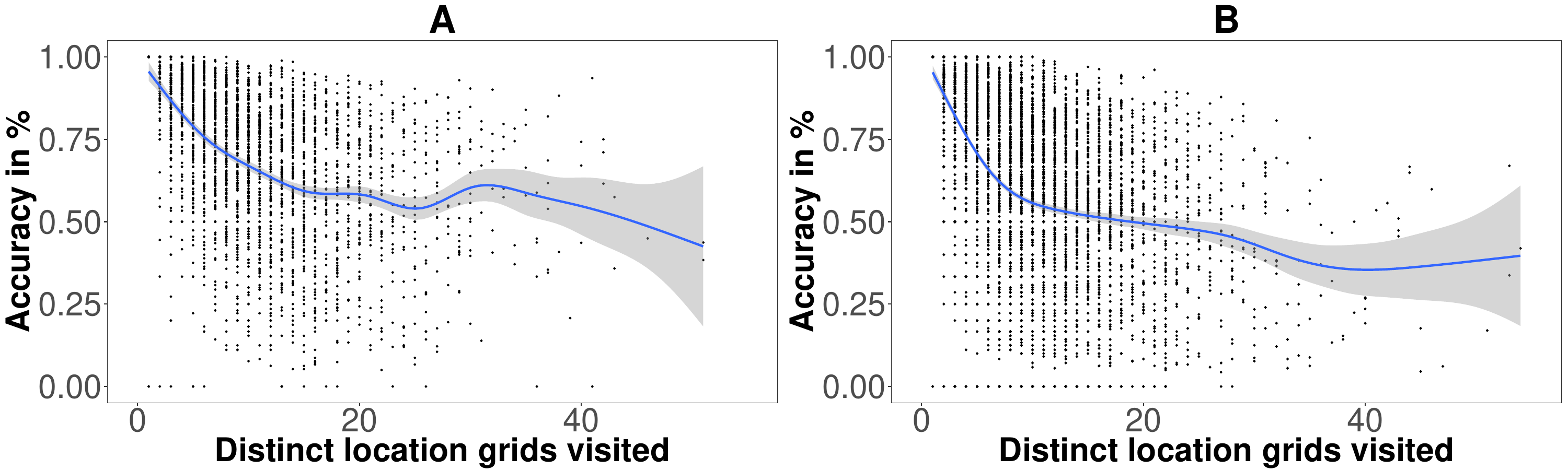}
\caption{A: Prediction accuracy (\%) for Top 1 vs. number of distinct locations visited for $r_i =1$ (A) and 2 (B).}
\label{distinct_loc_accuracy}
\end{figure}

Fig. \ref{distinct_loc_accuracy} shows prediction accuracy versus the number of distinct locations grids by individuals, for 1, and 2-hour resolutions; accuracy decreases with an increase in the number of distinct location grids visited by an individual. The fitted line is generated using a generalized additive model (GAM).

\section{Conclusion}
In this paper, we present a method for predicting missing locations from an individual's mobility timeline with good accuracy, using only sparse location tags from social media data. 
In order to address the challenge of sparse data, the model uses several heuristics of human movement and incorporates similar user data.
The proposed approach consistently outperforms baseline heuristic based methods across data from three major cities, showing stability of the approach. We also show how ILC fulfills timeline prediction better than an RNN in sparse data settings, though use of heuristics should be incorporated into the RNN architecture design in future work to further advance the approach.

We recognize limitations of this work. Predictions for an individual can be biased based on their Tweeting patterns (which can be specific to the types of people who use Twitter), although the incorporation of community behavior helps minimize this bias. Second, even though our work advances previous work by predicting full timelines for a large number of users, there are still many users for whom the location cannot be predicted by our model. Hence, despite the generalizability of the method and the dataset, the methodology will not be accurate for every single user. Third, here ILC only uses one location point in the past i.e $k-1$ to predict the location at $k$ due to the sparse nature of the data and prioritization of filling in the timeline, but we can expand the approach to use the sequence of $n$ locations in the past to predict the next location, with more complex considerations. Overall, this research demonstrates a new approach for the specific problem of filling in location timelines from sparse social media data, without assuming any information besides location data is available. The result can be used in many real-world applications that require location timelines.

\begin{acks}
Support for this project was provided in part by a grant from the National Science Foundation (1737987). We acknowledge Prof. Juliana Freire of New York University and her group for assistance with data.
\end{acks}

\bibliographystyle{ACM-Reference-Format}

\bibliography{references_lib.bib}

\end{document}